\documentclass[a4paper,11pt]{article}
\usepackage[utf8]{inputenc}
\usepackage{multicol,caption}
\usepackage{lipsum}
\usepackage[margin=.75in]{geometry}
\usepackage{graphicx}
\usepackage{amsmath}
\usepackage{amssymb}
\usepackage{float}
\usepackage{braket}

\newcommand{\M}{\mathcal{M}}
\newcommand{\dM}{\mathcal{\partial M}}
\newcommand{\bA}{\overline{A}}

\title{Holographic Entanglement Entropy and the $(3+1)$-dimensional Topological Black Hole}
\author{Zachary Polonsky\footnote{polonsza@mail.uc.edu}, Alex Flournoy\footnote{aflourno@mines.edu}\\
\\
\textit{Department of Physics, Colorado School of Mines, Golden, CO 80401, USA}}
\date{}

\begin{document}

\maketitle
\begin{abstract}
We investigate the Holographic Entanglement Entropy proposal in the context of the $(3+1)$-dimensional topological black hole. In contrast to the well-studied $(2+1)$-dimensional case, the maximal extension for this black hole includes only a single exterior region with its conformal boundary. This immediately raises a puzzle as to how one can view the purification of the dual conformal field theory state in terms of a thermofield double in the usual manner. Motivated by this puzzle, we calculate the horizon area for these black holes and discover that the result is observer dependent. This observer dependence poses a potential issue in applying the holographic entropy proposal. Investigating this we find that, although this observer dependence does not carry over to the holographic entanglement entropy, there is an indication of a coordinate system which is best adapted for the holographic calculation. These coordinates only cover two regions of the spacetime which exactly correspond to the regions of the CFT on which particle modes are well defined and so we see that the holographic calculation in the spacetime is capable of predicting regions of the CFT where particles cannot exist.
\end{abstract}
\newpage
\noindent\rule{\textwidth}{0.4pt}
\tableofcontents
\noindent\rule{\textwidth}{0.4pt}
\section{Introduction}

In condensed matter systems and general quantum field theories, it is important to know the number of degrees of freedom which contribute to the dynamical behavior of the system. One way of counting these degrees of freedom is by calculating the entanglement entropy of the system in question. However, directly calculating entanglement entropy in quantum field theories is difficult beyond certain 2-dimensional systems \cite{hrt,cardy}.

For the past two decades, the AdS/CFT-correspondence has provided a context for addressing calculations in conformal field theories with (often simpler) calculations in a dual gravitational theory \cite{maldacena1}. An incredibly useful realization of this comes in the form of Ryu and Takayanagi's Holographic Entanglement Entropy (HEE) proposal and its covariant generalization by Hubeny, Rangamani, and Takayanagi \cite{ryutak,hrt,aspects}. The HEE proposal asserts that the entanglement entropy in a CFT can be calculated from the area of an associated surface in the dual spacetime. 

Up to this point, most holographic calculations have been restricted to the relatively simple settings of spacetimes in $(2+1)$-dimensions and CFTs in $(1+1)$-dimensions which exhibit no time-dependence \cite{ryutak,hrt,aspects}. To use HEE for physical systems, the results in $(2+1)$-dimensional spacetimes and $(1+1)$-dimensional CFTs must be generalized to higher dimensions and time-dependent systems. This work provides some steps in this direction.

In the HEE proposal we consider an asymptotically AdS spacetime manifold, $\M$, and a CFT defined on the fixed boundary geometry, $\dM$. Splitting a spatial slice of the CFT into two regions, $A$ and $\bA$, such that $\dM=A\cup \bA$, the entanglement entropy of subregion $A$ will typically be given by

\begin{equation}
 S_A=Tr_A\left(\rho_A \log \rho_A\right) \quad , \quad \rho_A=Tr_{\bA} \rho
\end{equation}
where $\rho=\ket{\Psi}\bra{\Psi}$ is the density matrix describing the state of the full CFT \cite{entanglement}. The act of tracing over all possible states in $\bA$ is equivalent to the statement that a measurement of $A$ carries no information of the state of $\bA$.

\begin{figure}
 \centering
 \includegraphics[width=.3\textwidth]{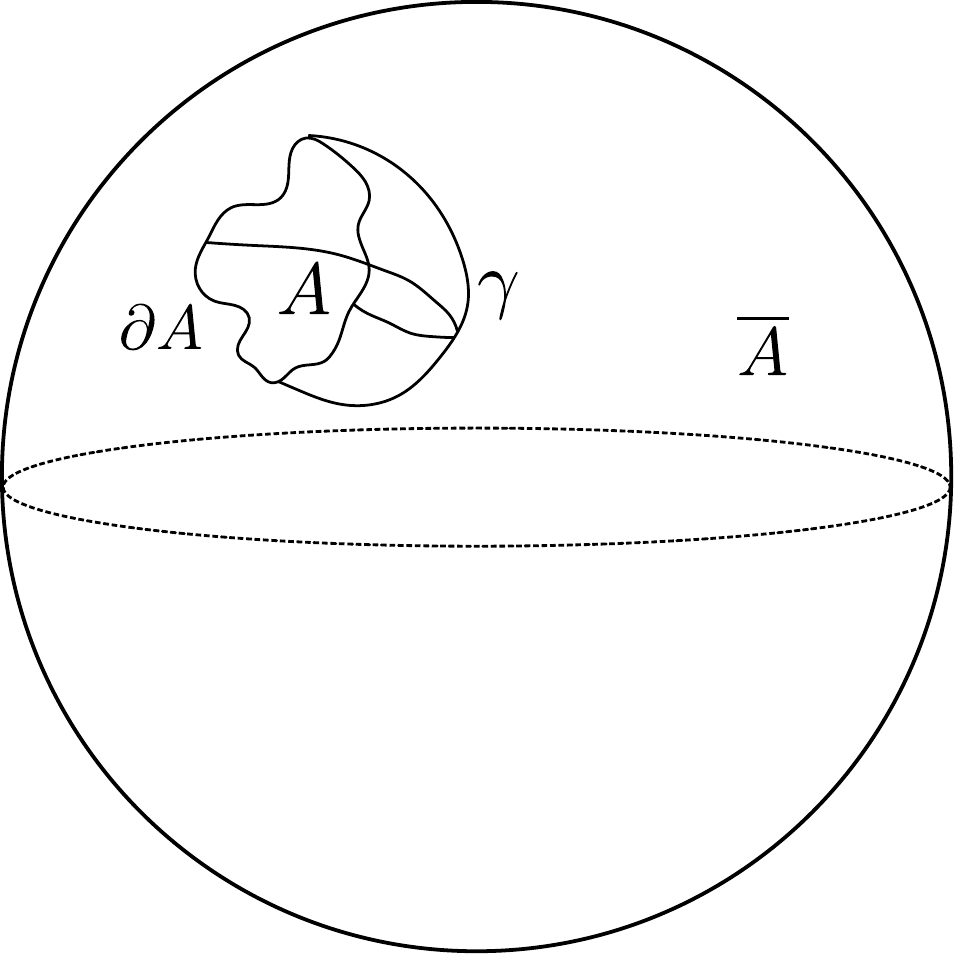}
 \captionof{figure}{The entanglement entropy of a spatial region $A$ of a CFT defined on the boundary of asymptotically AdS spacetime can be given by calculating the area of the minimal-area surface, $\gamma$, in the bulk which terminates on the boundary of $A$, $\partial A$.}
 \label{hee}
\end{figure}

We can mirror this idea in the dual spacetime by requiring that an observer with access to the region $A$ of $\dM$ cannot gain any information from the region $\bA$ of $\dM$. However, this is the same as requiring $\bA$ to be behind a horizon which terminates on the boundary of $A$, $\partial A$. From the lessons of Bekenstein and Hawking, we know that the entropy of a horizon is given by

\begin{equation}
 S_{H}=\frac{\mathrm{Area(horizon)}}{4G_N^{(d)}}
\end{equation}
where $G_N^{(d)}$ is the $d-$dimensional gravitational constant \cite{area,bhent,beken}. However, there is a continuous family of surfaces in $\M$ which terminate on $\partial A$, so we choose the unique surface with minimal area, $\gamma$. Then the HEE proposal states that the entanglement entropy of $A$ is given by

\begin{equation}
 S_A=\frac{\mathrm{Area}(\gamma)}{4G_N^{(d)}}.
\end{equation}
A pictorial representation of HEE can be seen in Figure \ref{hee}. Results of HEE have been compared to direct calculations in CFTs using the Cardy formula and show exact agreement \cite{ryutak,hrt,aspects}.

As is common with accelerated observers in flat spacetimes, different observers can see different horizon areas. These disagreements generally do not cause conflicts with HEE due to the fact that, as in the accelerated case, introducing these horizons typically cuts observers off from the full boundary \cite{rindlerads}. However, as we will discover, there exist spacetimes where different observers with access to the full boundary disagree on horizon areas. This introduces the possibility for a ``correct'' coordinate system in which to perform the HEE calculation. 

In this paper we will begin by reviewing the construction of the $(3+1)$-dimensional topological black hole as well as three coordinate systems that will be of later use in exploring its properties. We compute the horizon area in each set of coordinates and identify discrepancies and important features of the results. We then apply the HEE proposal to evaluate the entanglement entropy of a region of the boundary theory in the various coordinate systems, obtaining the important and expected result that the entropy of a given region on the boundary is independent of the coordinates used. We do however identify one of these as better adapted to the calculation. In order to understand these results, we examine the particle modes on the boundary theory and identify that the certain restrictions on the coordinates in the bulk reflect the restricted regions over which particle modes on the boundary are defined. Moreover we obtain a better understanding of why one of the three coordinate systems is better adapted to the use of the HEE proposal. We end with some conclusions and outlook. 

\section{$(3+1)$-Dimensional Topological Black Hole}

\begin{figure}
 \centering
 \includegraphics[width=.5\textwidth]{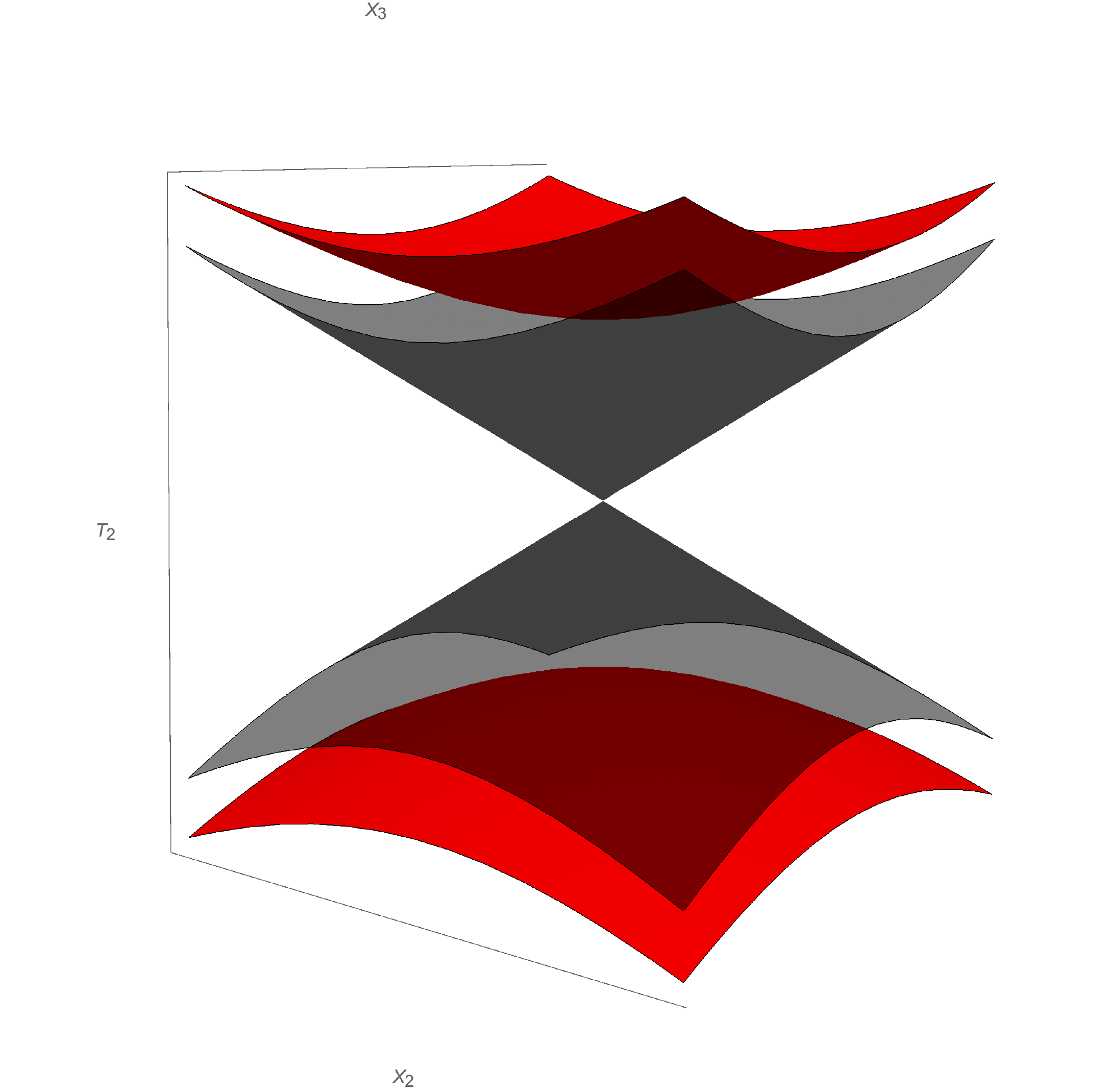}
 \captionof{figure}{Singularity (red) and event horizon (black) created from a quotient by $\xi$. Here, the $T_1$ and $X_1$ coordinates are suppressed.}
 \label{blackhole}
\end{figure}

The $(3+1)$-dimensional topological black hole can be formed as a quotient of global AdS, just like its $(2+1)$-dimensional analog \cite{btz,topbh,quo,brill,btzlong}. We begin with AdS$_4$, defined as the surface

\begin{equation}
\label{ads4}
 -T_1^2-T_2^2+X_1^2+X_2^2+X_3^2=-1
\end{equation}
embedded in $\mathbb{R}^{2,3}$ with metric

\begin{equation}
\label{r23}
 ds^2=-dT_1^2-dT_2^2+dX_1^2+dX_2^2+dX_3^2
\end{equation}
where we have set the AdS radius to unity. We will consider the boost-like isometry given by

\begin{equation}
\label{quovec}
 \xi=-X_1\partial_{T_1}-T_1\partial_{X_1} \quad , \quad \xi^2=-X_1^2+T_1^2
\end{equation}
to generate the quotient. To avoid closed timelike curves, we remove regions of the spacetime where $\xi^2<0$ after the quotient \cite{quo,brill}. This creates a singularity in the causal structure where timelike geodesics can end at

\begin{equation}
 -X_1^2+T_1^2=0
\end{equation}
or equivalently, by using (\ref{ads4}),

\begin{equation}
 -T_2^2+X_2^2+X_3^2=-1.
\end{equation}
This singularity asymptotes to the null cone given by

\begin{equation}
 T_2^2=X_2^2+X_3^2 \quad \Rightarrow \quad X_1^2-T_1^2=-1
\end{equation}
which we identify as the event horizon of the black hole. These surfaces are plotted in Figure \ref{blackhole}.

To examine the specific behavior of the $(3+1)$-dimensional topological black hole, we must define coordinates on the surface (\ref{ads4}). We will consider three coordinate systems adapted to different observers which will be important for studying how observational differences translate to HEE calculations.

The first set of coordinates are adapted to observers falling into the black hole and are related to the coordinates of the embedding space by

\begin{align}
 T_1=\frac{1-t^2+y_1^2+y_2^2}{1+t^2-y_1^2-y_2^2}&\cosh \phi \quad , \quad
 T_2=\frac{2t}{1+t^2-y_1^2-y_2^2} \nonumber \\
 \nonumber \\
 X_1=\frac{1-t^2+y_1^2+y_2^2}{1+t^2-y_1^2-y_2^2}\sinh \phi  \quad , \quad
 X_2&=\frac{2y_1}{1+t^2-y_1^2-y_2^2} \quad , \quad
 X_3=\frac{2y_2}{1+t^2-y_1^2-y_2^2}. 
\end{align}
We will refer to these as Kruskal coordinates for reasons to be pointed out a bit later. In terms of these coordinates the induced metric on the surface becomes

\begin{align}
\label{carkrusk}
 ds^2=\frac{4}{(1+t^2-y_1^2-y_2^2)^2}(-dt^2+dy_1^2+dy_2^2)
 +\left(\frac{1-t^2+y_1^2+y_2^2}{1+t^2-y_1^2-y_2^2}\right)^2d\phi^2
\end{align}
and the Killing vector which generates the quotient becomes $\xi=\partial_\phi$ \cite{topbh}. The quotient makes the identification $\phi=\phi+2\pi$, and the coordinate ranges after the quotient are given by $t,y_i\in(-\infty,\infty)$ and $\phi\in[0,2\pi)$. In these coordinates the singularity is given by the surface

\begin{align}
 \xi^2=\left(\frac{1-t^2+y_1^2+y_2^2}{1+t^2-y_1^2-y_2^2}\right)^2=0\; \rightarrow \; -t^2+y_1^2+y_2^2=-1
\end{align}
and the event horizon is given by

\begin{align}
 T_1^2-X_1^2=\left(\frac{1-t^2+y_1^2+y_2^2}{1+t^2-y_1^2-y_2^2}\right)^2=1 \; \rightarrow \; -t^2+y_1^2+y_2^2=0.
\end{align}

We can also see that the embedding coordinates diverge at the surface

\begin{equation}
 1+t^2-y_1^2-y_2^2=0
\end{equation}

which we associate with the boundary of the spacetime. Here, we note that, aside from the singularity, the spacetime in these coordinates is geodesically complete, and therefore no further maximal extension is necessary, hence the appropriateness of "Kruskal". In addition we see that this spacetime has only a single, connected asymptotic boundary \cite{topbh}.

For later calculations, it will be convenient to define the polar form of the Kruskal coordinates by $y_1=\chi \cos \theta$ and $y_2=\chi \sin \theta$ so that the induced metric becomes

\begin{align}
\label{polkrusk}
 ds^2=\frac{4}{(1+t^2-\chi^2)^2}(-dt^2+d\chi^2+\chi^2d\theta^2)
 +\left(\frac{1-t^2+\chi^2}{1+t^2-\chi^2}\right)^2d\phi^2
\end{align}
with $\chi\in[0,\infty)$ and $\theta\in[0,2\pi)$. Here, the singularity is given by $t^2-\chi^2=1$, the event horizon is given by $t^2-\chi^2=0$ and the boundary is given by $t^2-\chi^2=-1$. 

The second set of coordinates we will use are related to the cartesian Kruskal coordinates by

\begin{align}
 t&=\rho\sinh\tau \nonumber \\
 y_1&=\rho\cos\theta\cosh\tau \nonumber \\
 y_2&=\rho\sin\theta\cosh\tau
\end{align}
with ranges $\tau\in(-\infty,\infty)$, $\rho\in[0,1)$, and $\theta\in[0,2\pi)$. The metric becomes \cite{topbh}

\begin{align}
\label{fullext}
 ds^2=\frac{4}{(1-\rho^2)^2}\left(-\rho^2d\tau^2+d\rho^2\right.+\left.\rho^2\cosh^2\tau d\theta^2\right) +\left(\frac{1+\rho^2}{1-\rho^2}\right)^2d\phi^2.
\end{align}

In these coordinates, the boundary is located at $\rho=1$ and the event horizon is located at $\rho=0$. However, since the coordinates only cover the region $0\le \rho < 1$, they are restricted to the exterior of the black hole. For this reason, we will refer to these as full exterior coordinates.

The final set of coordinates we will consider are related to the cartesian Kruskal coordinates by

\begin{align}
 t=&\rho\sin\psi\sinh\zeta \nonumber \\
 y_1=&\rho \sin \psi \cosh \zeta \nonumber \\
 y_2=&\rho \cos \psi
\end{align}
with $\zeta\in(-\infty,\infty)$, $\rho\in[0,1)$, and $\psi\in[0,\pi]$ \cite{topbh}. This takes the metric to

\begin{align}
\label{static}
 ds^2=\frac{4}{(1-\rho^2)^2}\left(-\rho^2\sin^2\psi\,d\zeta^2\right.\left.+d\rho^2+\rho^2d\psi^2\right)  +\left(\frac{1+\rho^2}{1-\rho^2}\right)^2 d\phi^2
\end{align}
where again, the boundary is at $\rho=1$ and the event horizon is at $\rho=0$. Like the full exterior coordinates, the coordinates (\ref{static}) are only well-defined for the exterior of the black hole, however they are also restricted to the region

\begin{equation}
 y_1^2-t^2\ge0.
\end{equation}
Although these coordinates do not cover even the full exterior of the black hole, they have the advantage of being static. For this reason, we will refer to these as static coordinates.

Since the geometries described by both the Kruskal and full exterior coordinates have access to the full boundary of the spacetime, they should both be suitable duals to the full CFT \cite{maldacena1,eternalbh}. However, as we will see in the next section, observers in Kruskal coordinates measure a different event horizon area from observers in full exterior coordinates. Our goal is to examine how this disagreement between valid choices of coordinates in the bulk is reflected in the boundary CFT.

\section{Observer-Dependent Horizon Area}

The thermodynamics of spacetimes are generally deeply tied to the area of horizons in the spacetime \cite{area,bhent,beken}. In polar Kruskal coordinates the event horizon is located at the surface $\chi=t$. The induced metric on this surface is given by
\begin{equation}
\label{imetk}
 d\sigma^2_{\chi=t}=4t^2d\theta^2+d\phi^2.
\end{equation}
The area of the event horizon is then given by

\begin{equation}
 \mathrm{Area}(\chi=t)=2|t|\int_0^{2\pi} d\theta \int_0^{2\pi}d\phi=8\pi^2|t|
\end{equation}
which exhibits a time-dependence.

If we instead examine the event horizon in full exterior coordinates or static coordinates, where the event horizon is located at $\rho=0$, we find the induced metric on the horizon to be

\begin{equation}
 d\sigma_{\rho=0}^2=d\phi^2.
\end{equation}
However, since this is the metric of a 1-dimensional surface, it necessarily has zero area in a $(3+1)$-dimensional spacetime.

This poses a puzzle when considering the application of HEE. In applying HEE the entanglement entropy of a region $A$ of the boundary CFT has contributions associated with the area of event horizons in the bulk \cite{ryutak,hrt,aspects}. If observers with access to the full boundary cannot agree on the area of the horizon, we may then expect different results from the holographic calculation. However, if the region $A$ does not change when we change coordinate systems, there is no reason to expect a different entanglement entropy. To investigate this seeming paradox, we will apply the HEE calculation for each coordinate system directly.

\section{Holographic Calculation}

Since the spacetime in Kruskal and full exterior coordinates is time-dependent, the HEE calculation would generally require the covariant proposal of Hubeny-Rangamani-Takayanagi \cite{hrt}. However, since the topological black hole is formed from identifications of global AdS, we can perform the calculation by simply relating the Kruskal and full exterior coordinates to Poincar\'{e} coordinates \cite{ryutak,hrt,aspects}. The metric of global AdS$_4$ in Poincar\'{e} coordinates is given by

\begin{equation}
 ds^2=\frac{1}{Z^2}\left(dW_+dW_-+dY^2+dZ^2\right)
\end{equation}
where $W_{\pm},Y\in(-\infty,\infty)$ and $Z\in(0,\infty)$ \cite{poincare}. Here, we are using the null form of Poincar\'{e} coordinates, which can be written as $W_\pm=X\pm T$ where $X$ and $T$ are spacelike and timelike coordinates, respectively. The boundary of the spacetime is at $Z=0$. If we consider a strip on the boundary at fixed time given by

\begin{equation}
 W_+-W_-=const \; \Rightarrow \; \Delta W_+-\Delta W_-=0,
\end{equation}
the length of the strip will be given by $\Delta Y=L$ and the width will be given by

\begin{equation}
\label{rsq}
 R^2=\Delta W_+ \Delta W_-.
\end{equation}

\begin{figure}
 \centering
 \includegraphics[width=.25\textwidth]{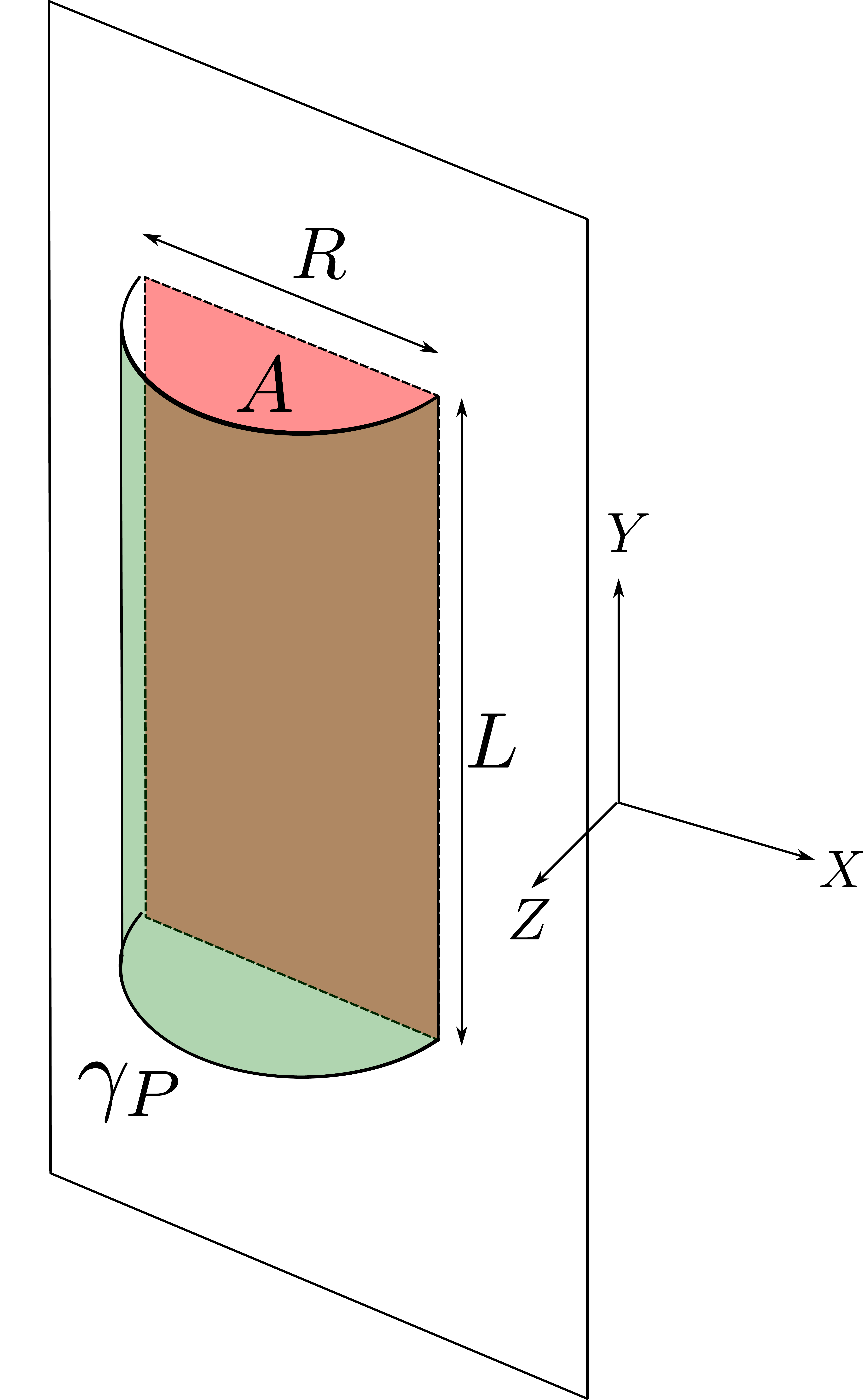}
 \captionof{figure}{Minimal-area surface, $\gamma_P$, (green) corresponding to the strip, $A$, (red) at a constant-time slice of global AdS$_4$.}
 \label{strip}
\end{figure}

If we choose a minimal-area surface, $\gamma_P$, which terminates only on the boundary in the $X=1/2(W_++W_-)$ direction as shown in Figure \ref{strip}, the area of $\gamma_P$ will be given by

\begin{equation}
\label{poinarea}
 \mathrm{Area}(\gamma_P)=2\left(\frac{L}{\epsilon}\right)-\alpha\left(\frac{L}{R}\right)
\end{equation}
where $\epsilon$ is a cutoff introduced to prevent the expression from diverging and

\begin{equation}
\alpha=4\pi \left(\frac{\Gamma\left(\frac{3}{4}\right)}{\Gamma\left(\frac{1}{4}\right)}\right)^2
\end{equation}
is a positive constant \cite{hrt,aspects}.

We will first find the area of the minimal-area surface in polar Kruskal coordinates which are related to Poincar\'{e} coordinates by

\begin{align}
 W_{\pm}&=\frac{2}{1-t^2+\chi^2}\left(\chi \cos\theta  \pm t\right)\,e^\phi \nonumber \\
Y&= \frac{2}{1-t^2+\chi^2}\; \chi\; \sin \theta \, e^\phi \nonumber \\
Z&=\frac{1+t^2-\chi^2}{1-t^2+\chi^2}\,e^\phi.
\end{align}

In these coordinates, the boundary is the surface where $Z=0$, which corresponds to $\chi=\sqrt{1+t^2}$. If we anchor the minimal-area surface, $\gamma$, to the boundary at fixed time $t=t_0$ such that $\theta\in[\theta_1,\theta_2]$ and $\phi\in[\phi_1,\phi_2]$, the cutoff at $Z=\epsilon\ll1$ corresponds to

\begin{equation}
\label{kep}
 \epsilon_{1,2}=ae^{\phi_{1,2}} \quad \mathrm{where} \quad 0<a\ll1.
\end{equation}
For simplicity, we will center the region $A$ on the boundary such that $\theta_1=-\Theta$ and $\theta_2=\Theta$. Thus, the length, $L$, of $A$ is given by

\begin{equation}
\label{kl}
 L=\Delta Y=\sqrt{1+t_0^2}\left(e^{\phi_2}+e^{\phi_1}\right)\sin\Theta
\end{equation}
and the width, $R$, is given by (\ref{rsq})

\begin{equation}
\label{krsq}
 R^2=\left(e^{\phi_2}-e^{\phi_1}\right)^2\left[\left(1+t_0^2\right)\cos^2\Theta-t_0^2\right].
\end{equation}

Using (\ref{kep}), (\ref{kl}), and (\ref{krsq}) in (\ref{poinarea}), we find an expression for the area of $\gamma$ using $\epsilon^2=\epsilon_1 \epsilon_2$

\begin{align}
 \label{areakrusk}
 \mathrm{Area}(\gamma)=\,4\left(\frac{\cosh \Delta}{a}\right)\sqrt{1+t_0^2}\sin \Theta-\alpha\left(\frac{\sqrt{1+t_0}\sin\Theta}{\sqrt{(1+t_0^2)\cos^2\Theta-t_0^2}}\right)\coth \Delta
\end{align}
where $\Delta=\left(\phi_2-\phi_1\right)/2$. From this expression, we can see that the surface, $\gamma$, is restricted to the region

\begin{equation}
\label{region1}
 \left(1+t_0^2\right)\cos^2\Theta-t_0^2>0
\end{equation}
where the expression (\ref{areakrusk}) is real.

Similarly, we can perform the same HEE calculation with full exterior coordinates, which are related to Poincar\'{e} coordinates by

\begin{align}
 W_{\pm}&=\frac{2\rho}{1+\rho^2}\left(\cosh\tau\cos\theta  \pm \sinh\tau \right)\,e^\phi \nonumber \\
Y&= \frac{2\rho}{1+\rho^2} \cosh \tau \sin \theta \, e^\phi \nonumber \\
Z&=\frac{1-\rho^2}{1+\rho^2}\,e^\phi.
\end{align}

For fixed time, $\tau=\tau_0$, the region $A$ on the boundary will again be given by the ranges $\theta\in[-\Theta,\Theta]$ and $\phi\in[\phi_1,\phi_2]$. We find the cutoff is given by

\begin{equation}
 \label{fep}
 \epsilon_{1,2}=ae^{\phi_{1,2}},
\end{equation}
the length of $A$ is given by

\begin{equation}
 L=\cosh\tau_0\left(e^{\phi_2}+e^{\phi_1}\right)\sin\Theta
\end{equation}
and the width is

\begin{equation}
R^2=(e^{\phi_2}-e^{\phi_1})^2(\cosh^2\tau_0\cos^2\Theta-\sinh^2\tau_0).
\end{equation}

Using these values in (\ref{poinarea}) gives the area of the surface, $\gamma$, in full exterior coordinates

\begin{align}
 \label{areafe}
 \mathrm{Area}(\gamma)=\,4\left(\frac{\cosh \Delta}{a}\right)\cosh\tau_0\sin \Theta-\alpha\left(\frac{\cosh\tau_0\sin\Theta}{\sqrt{\cosh^2\tau_0\cos^2\Theta-\sinh^2\tau_0}}\right)\coth \Delta.
\end{align}

If we consider Kruskal and full exterior coordinates, we can see that $\cosh\tau_0=\sqrt{1+t_0^2}$ and so both coordinates give the same result for the entanglement entropy of the region $A$ of the boundary CFT. 

Like before, the area of the surface $\gamma$ is not real over the full boundary and is restricted to the region

\begin{equation}
\label{region2}
 \cosh^2\tau_0\,\cos^2\Theta-\sinh^2\tau_0>0
\end{equation}
which is the same as the region (\ref{region1}). We recall that static coordinates are confined to

\begin{equation}
y_1^2-t^2>0 \; \to \; \chi^2\cos^2\theta-t^2>0  
\end{equation}
which, on the boundary at fixed time given by $t=t_0$, $\chi=\sqrt{1+t_0^2}$, and $\theta=\Theta$, corresponds exactly to (\ref{region1}) and (\ref{region2}). Thus, it seems that static coordinates are naturally adapted to the HEE calculation.

As we may have expected, the entanglement entropy of the CFT on region $A$ of the boundary does not vary between coordinate systems using the holographic calculation. However, we have arrived at the somewhat surprising result that there is a preferred coordinate system for the HEE calculation. Furthermore, the coordinate system which is preferred does not cover the full boundary.

\section{Restriction of Particle Modes}

To better understand the results of the HEE calculation, we can turn to the theory on the boundary. First, we will define global coordinates on AdS$_4$ in terms of the embedding coordinates by

\begin{align}
 T_1=\frac{1+r^2}{1-r^2}&\cos t\quad , \quad 
 T_2=\frac{1+r^2}{1-r^2}\sin t \nonumber \\
 \nonumber \\
 X_1=\frac{2r}{1-r^2}\cos\lambda \quad , \quad 
 X_2&=\frac{2r}{1-r^2}\cos\theta\,\sin\lambda \quad , \quad
 X_3=\frac{2r}{1-r^2}\sin\theta\,\sin\lambda
\end{align}
with ranges $r\in[0,1)$, $\theta\in[0,2\pi)$, $\lambda\in[0,\pi]$, and $t\in(-\infty,\infty)$ after enforcing a universal covering to avoid closed timelike curves \cite{ads}. The metric becomes

\begin{align}
\label{gloads}
 ds^2=\frac{4}{\left(1-r^2\right)^2}\left[-\frac{\left(1+r^2\right)^2}{4}\,dt^2 +dr^2+r^2d\lambda^2+r^2\cos^2\lambda\, d\theta^2\right]
\end{align}
where the boundary is located at $r=1$. The Killing vector that generates the quotient to form the black hole is given by (\ref{quovec}). There is another Killing vector orthogonal to $\xi$ given by $\eta=X_2\partial_{T_2}+T_2\partial_{X_2}$, where the motivation for introducing $\eta$ will become clear. To get the metric on the boundary, we multiply (\ref{gloads}) by the conformal factor

\begin{equation}
 \Omega^2=\frac{\left(1-r^2\right)^2}{4}
\end{equation}
and take $r\to1$ to obtain

\begin{equation}
 d\sigma^2=-dt^2+d\lambda^2+\cos^2\lambda\,d\theta^2.
\end{equation}

The conformal Killing vectors on the boundary corresponding to $\xi$ and $\eta$ are given by

\begin{align}
 \xi_b=\cos\lambda\,\sin t\,\partial_t+\cos t\,&\sin\lambda\,\partial_\lambda \nonumber \\
 \eta_b=\cos t\,\sin\lambda\,\cos\theta\,\partial_t+&\sin t\,\cos \lambda \,\cos \theta \partial_\lambda-\sin t\,\csc \lambda \,\sin\theta \, \partial_\theta
\end{align}
respectively \cite{ckv}. To find the region of the boundary that will survive the quotient, we must find the region where $\xi_b^2>0$. This region corresponds to the diamond

\begin{equation}
 \left\{(\lambda,t)\left|0<\lambda<\pi, \; |t|<\frac{\pi}{2}-\left|\lambda-\frac{\pi}{2}\right|\right.\right\}
\end{equation}
in contrast to the BTZ black hole, which gives two separate diamonds \cite{loukomarolf,louko}. This is a reflection of the fact that the $(3+1)$-dimensional topological black hole has a single, connected boundary.

To better understand the action of $\xi_b$ on the boundary, we define the coordinates (inspired by those used in \cite{loukomarolf})

\begin{align}
 \alpha&=-\log\left[\tan\left(\frac{\lambda-t}{2}\right)\right] \nonumber \\
 \beta&=\log\left[\tan\left(\frac{\lambda+t}{2}\right)\right]
\end{align}
giving the metric

\begin{equation}
\label{notinv}
 d\sigma^2=\frac{d\alpha d\beta +\cosh^2\left[(\alpha+\beta)/2\right]\,d\theta^2}{\cosh\alpha\,\cosh\beta}
\end{equation}
where the conformal Killing vectors are now given by

\begin{align}
 \xi_b&=-\partial_\alpha+\partial_\beta \nonumber \\
 \eta_b&=\cos\theta\,(\partial_\alpha+\partial_\beta)-\sin\theta\tanh\left(\frac{\alpha+\beta}{2}\right)\partial_\theta.
\end{align}
The vector $\xi_b$ maps the point 

\begin{equation}
(\alpha,\beta,\theta)\to(\alpha-c,\beta+c,\theta)
\end{equation}
where $c$ is a constant. Unfortunately, the metric (\ref{notinv}) is not invariant under an action generated by $\xi_b$, but the conformally related metric

\begin{equation}
 d\sigma^2=d\alpha\,d\beta+\cosh^2\left(\frac{\alpha+\beta}{2}\right)d\theta^2
\end{equation}
is. Finally, defining $\alpha=\tau-\phi$ and $\beta=\tau+\phi$ where $\tau,\phi\in(-\infty,\infty)$, we obtain the metric

\begin{equation}
\label{bound}
 d\sigma^2=-d\tau^2+\cosh^2\tau\,d\theta^2+d\phi^2
\end{equation}
and the conformal Killing vectors take the form

\begin{align}
 \xi_b&=\partial_\phi \nonumber \\
 \eta_b&=\cos\theta\,\partial_\tau-\sin\theta\,\tanh\tau \, \partial_\theta.
\end{align}
A quotient by $\xi_b$ makes the identification $\phi=\phi+2\pi$ and we see that the metric (\ref{bound}) is the conformal boundary of the topological black hole in full exterior coordinates (\ref{fullext}).

\begin{figure}
 \centering
 \includegraphics[width=.5\textwidth]{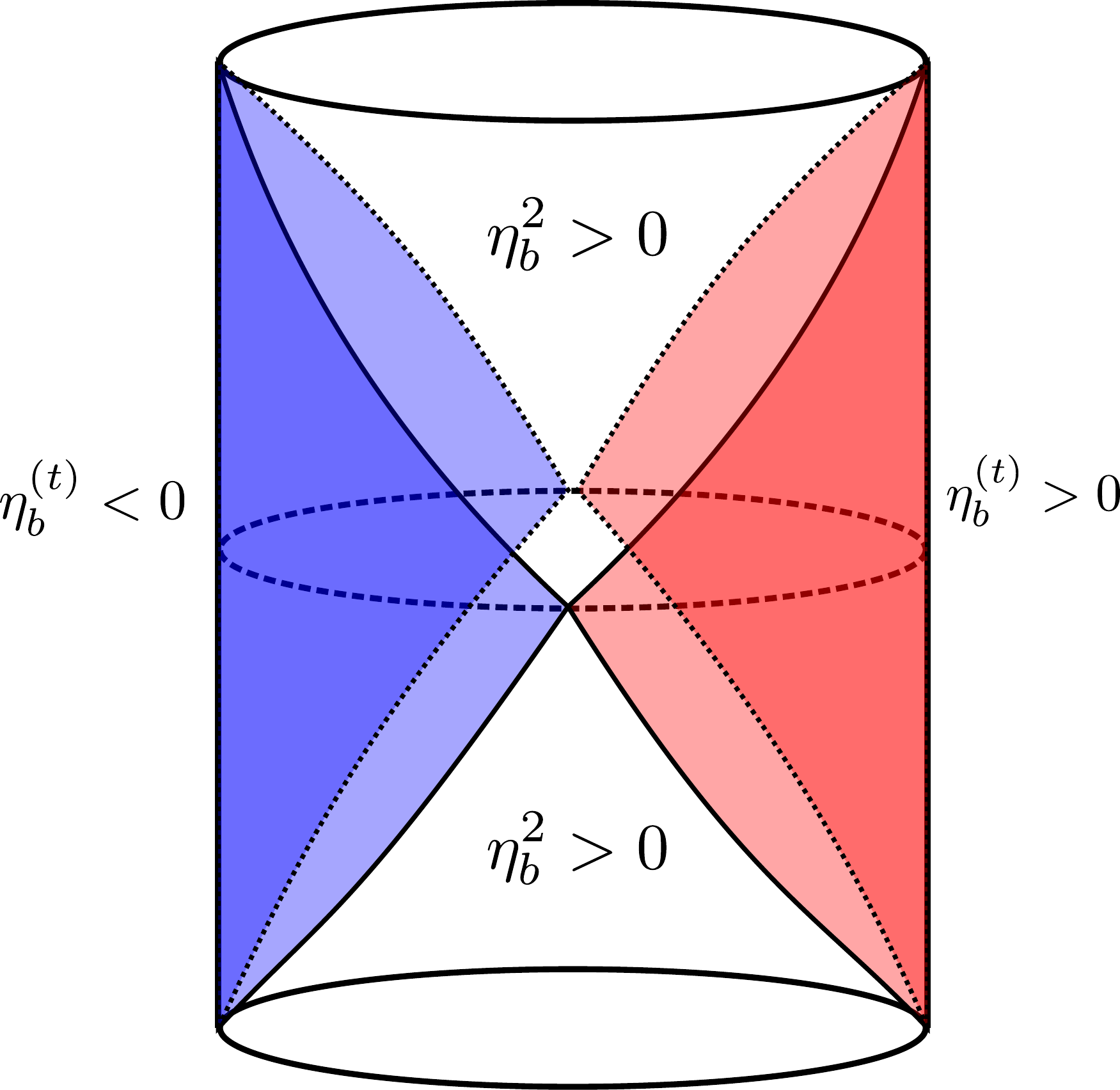}
 \captionof{figure}{Diagram of the boundary of the topological black hole. The dashed circle represents $t=\tau=0$, where $\eta_b$ is timelike over the whole boundary. We see two distinct regions on the boundary where $\eta_b$ is timelike, one where the time component, $\eta_b^{(t)}$, is positive, which we associate with positive energy modes (red), while on the other, $\eta_b^{(t)}$ is negative, which we associate with negative energy modes (blue).}
 \label{tbhstate}
\end{figure}

Near $\tau=0$, we find $\eta_b=\cos\theta\, \partial_\tau$ is purely timelike, except for the points $\theta=\pm\pi/2$. Moreover, for the region

\begin{equation}
 D_f\doteq\,\left\{\theta\in\left(-\frac{\pi}{2},\frac{\pi}{2}\right)\right\}
\end{equation}
$\eta_b$ is future-directed, while for the region 

\begin{equation}
 D_p\doteq\, \left\{\theta\in\left(\frac{\pi}{2},\frac{3\pi}{2}\right)\right\}
\end{equation}
$\eta_b$ is past-directed. Therefore, we naturally associate $\eta_b$ with particle modes in the boundary theory, where positive energy modes are associated with $\eta_b$ in $D_f$ and negative energy modes are associated with $\eta_b$ in $D_p$ (Figure \ref{tbhstate}).

As the system evolves away from $\tau=0$, we find that $\eta_b$ is no longer timelike over the full boundary and thus particles are not well defined on the regions where $\eta_b^2>0$ given by

\begin{equation}
 \cosh^2\tau\,\cos^2\theta-\sinh^2\tau>0.
\end{equation}
However, this is exactly the region of the boundary to which the surface $\gamma$ is restricted in the holographic calculation (\ref{region2}). This is a result that could have been anticipated, since we should not expect to have contributions to entanglement entropy from regions of the CFT were particles are disallowed.

Here, we see the reason why the HEE calculation has resulted in a surface which is not well-defined over the full boundary of the spacetime: we can only have entanglement in regions where we can have particles. Since we associate particle modes with timelike Killing vectors, it makes sense that the coordinate system best adapted to the HEE calculation is that which has a global timelike Killing vector, i.e. static coordinates in the case of the $(3+1)$-dimensional topological black hole.

\section{Conclusions}

The case of the $(3+1)$-dimensional topological black hole is of interest due to the fact that it is a time-dependent spacetime with an observer-dependent event horizon. While horizons which vary between observers appear elsewhere in relativity, this particular case allows two observers with access to the full boundary to see different event horizon areas. In previous holographic entanglement entropy calculations, it has been shown that entanglement of the CFT on the boundary is closely related to the area of event horizons in the bulk. This suggests the possibility of an observer-dependence in holographic entanglement entropy.

While our calculation of HEE shows no observer-dependence, it does suggest that there is a preferred coordinate system for the calculation, i.e. static coordinates, which is somewhat surprising given that the coordinates do not cover the full spacetime. Moreover, these coordinates are such that the metric is time-independent, i.e. has a global timelike Killing vector. Since the region of the spacetime which these coordinates cover correspond exactly to the region in the boundary theory where the Killing vector associated with particle modes is timelike, we expect this to hold for other spacetimes as well. This is a powerful result, since the time-independent holographic entanglement entropy calculation of Ryu-Takayanagi is significantly less involved than the covariant prescription of Hubeny-Rangamani-Takayanagi. This result also shows that, while holographic entanglement entropy can give insight to the properties of the dual CFT, it can also predict the regions in the CFT where particles are disallowed.

There is further work to be done by generalizing this work to other time-dependent spacetimes and their corresponding CFT duals, as well as generalizing these results to even higher dimensions. The challenge of these generalizations is that most time-dependent spacetimes do not exhibit the convenient property of being locally AdS, and therefore the methods used in this paper cannot be applied. Therefore, to properly generalize these results, one must use the covariant method in these cases. Luckily, there are still many interesting spacetimes to study which can be obtained from a single quotient or multiple quotients of global AdS.

\centering{
\noindent\rule{8cm}{0.4pt}
}

\end{document}